\documentclass[prl,aps,twocolumn,superscriptaddress,showpacs]{revtex4}

\usepackage{epsfig}

\newcommand{\be}{\begin{equation}}
\newcommand{\ee}{\end{equation}}
\newcommand{\bea}{\begin{eqnarray}}
\newcommand{\eea}{\end{eqnarray}}

\newcommand{\la}{\langle}
\newcommand{\ra}{\rangle}

\newcommand{\ri}{\text{i}}
\newcommand{\re}{\text{e}}

\def\nn{\nonumber\\}

\begin{document}

\title{Spectrum and correlation functions of a quasi-one-dimensional quantum Ising model}

\author{Sam T. Carr}
\affiliation{Dept Physics, Building 510A, Brookhaven National Laboratory, Upton, NY 11973 }
\affiliation{Dept Theoretical Physics, 1 Keble Road, Oxford OX1 3NP}
\author{Alexei M. Tsvelik}
\affiliation{Dept Physics, Building 510A, Brookhaven National Laboratory, Upton, NY 11973 }

\begin{abstract}
For  a model of weakly coupled quantum  Ising chains we outline   the phase diagram  and establish that well below the transition line the system has a remarkably one dimensional spectrum. We study the dynamical magnetic susceptibility and find a very rich spectrum with several modes corresponding to a fundamental particle and its bound states.  The approach is based on  Bethe ansatz and  the Random Phase Approximation applied to the interchain exchange.  
\end{abstract}

\pacs{75.10.Pq, 75.40.Gb}

\maketitle


The Ising model is one of the generic models for many quantum and classical 
 systems.  Besides  extensive applications to spin systems  (here we refer the reader to \cite{qpt,cds96,mat85}), the model could also be
used to describe  interacting electric dipoles (like in systems with orbital degrees of freedom \cite{kk82}) or arrays of interacting Josephson junctions (see, for example, \cite{dfi02}). The low dimensions are the most interesting since here the model exhibits strongest correlations. In one dimension the quantum Ising model (see Eq.(\ref{qim}) below) is exactly solvable by means of Jordan-Wigner transformation which converts the spin Hamiltonian into a Hamiltonian of non-interacting fermions. In general the spectrum has a gap which is closed when the transverse magnetic field is equal to the exchange integral ($g =0$). The exact solution also exists when the  magnetic field has $z$-component, although only when the $x$-component is kept equal to $J$ \cite{zam88}. The solution predicts a rich spectrum with as many as eight particles and a hidden E$_8$ symmetry. An interesting question is whether some of this fascinating physics may survive in realistic systems which are almost never truly one-dimensional. The goal of this paper is to discuss this and related problems.  We begin by introducing the model, go on to outline  the phase diagram and then discuss  spin-spin correlation functions in the ordered phase.  In certain regions in parameter space we see that the one-dimensional behavior is clearly visible.

 Let us consider the quasi-one-dimensional (weakly coupled one-dimensional chains) quantum Ising model described by the following Hamiltonian:
\bea
H & = & \sum_j H_{1D}^{(j)} + \sum_{i,j,n,m} J_{nm}(i-j) 
 \sigma_i^z (n) \sigma_j^z (m) \label{ham}\\
H_{1D}^{(j)} & = & -J_\| \sum_n \left\{ \sigma^z_j (n) \sigma^z_j (n+1) + (1+g)\sigma^x_j (n) \right\},
\label{qim}
\eea
where $i,j$ label the chains, $n,m$ label the sites on the chain and $\sigma$ are the Pauli spin operators.
For  simplicity we discuss  the case where $J_{nm}(i-j) = J_\perp (i-j)$ if $n=m$ and $0$ otherwise.

\begin{figure}
\begin{center}
\epsfig{file=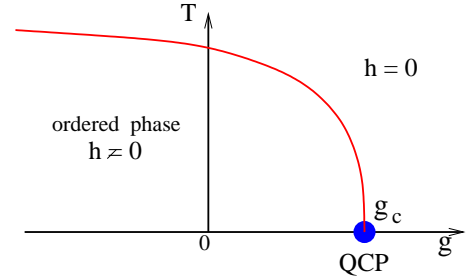,width=2.5in}
\caption{ Expected Phase Diagram of model (\ref{ham}). Estimates of  $g_c$ 
 and $T_c(g=0)$ are  given by  Eq.\ \ref{qcp1} and Eq.\ \ref{tcg0}  respectively. In the vicinity of $g =0$ point the physics is well described by model (\ref{mf}). }\label{phase1}
\end{center}
\end{figure}

 The dimensionless parameter $1+g$ is the strength of the transverse field written in a way such that the one-dimensional model has a phase transition at $T =0$ at $g =0$. At $T =0$ the phase with $g < 0$ is ordered with $\langle \sigma^z\rangle \neq 0$ and the phase with $g > 0$ is quantum disordered  (see e.g. \cite{qpt}). We would expect a non-zero inter-chain coupling to extend the ordered region to finite temperatures and shift the critical coupling to $g \neq 0$ as shown schematically in Fig. \ref{phase1}.

 The spectrum and dynamics of the model close to the transition line is very well understood \cite{id89a}. The transition itself falls into the 3D Ising model universality class, the Quantum Critical Point (QCP) falls into the 4D Ising model universality class. We concentrate our attention on the region of phase diagram well below the transition line where new non-universal physics can be found.  We demonstrate that non-trivial physical effects are possible in this region. 

 If we are not very close to the phase transition line, the inter-chain 
coupling can be taken into account in the Random Phase Approximation (RPA). The RPA expression for the dynamical magnetic susceptibility in the disordered phase is \be
\chi(\omega,k;{\bf k_\perp}) = [\chi_{1D}^{-1}(\omega,k) - J_\perp ({\bf k_\perp})]^{-1}.
\label{rpa}
\ee 
where $\chi_{1D}$ is the susceptibility of a single chain: 
\be
\chi_{1D}(\omega,k) = -\ri \sum_n \int_0^\infty dt
\re^{\ri \omega t - \ri k n} \langle \left[ \sigma^z (t,n), \sigma^z (0,0) \right]\rangle,
\label{ftd}
\ee
where  $v=J_\| a_0$ is the on-chain velocity and $a_0$ is the lattice spacing in the
chain direction. 
Formally RPA is the first term in an expansion in the inverse transverse coordination number $1/z_\perp$, where $z_\perp < 6$ if we are embedding our chains within a three dimensional structure. The application of RPA  to other quasi-one-dimensional magnets \cite{ik00,boc01} have demonstrated that far from the transition it provides a good accuracy even for small  coordination numbers. 

In the ordered state expression (\ref{ftd}) has to be modified. Namely, one has to replace $\chi_{1D}$ by the dynamical susceptibility calculated in the presence of an effective magnetic field generated by the neighboring chains \cite{sch96}. In other words, in calculating $\chi_{1D}$ one has to use the following Hamiltonian:
\be
H_{1D} = \sum_n\left\{ -J_\| \left[\sigma^z (n) \sigma^z (n+1) + (1 + g)\sigma^x (n)\right] + h\sigma^z (n)\right\}\label{mf}
\ee
with the self-consistency relation
\be
h = J_\perp (q=0)\langle \sigma^z \rangle.\label{mfsc}
\ee

There are three energy scales in this problem.  $J_\|$ the on chain coupling, $m=|g|J_\|$ is the spectral  gap for  a single chain, and $J_\perp$ the inter-chain coupling.  In order to treat the model as weakly coupled chains we must have $J_\perp\ll J_\|$ and in order to apply the non-perturbative results for the individual chains which have all been found in the continuous limit, we must have $m \ll J_\|$.  However, $m$ and $J_\perp$ can both be of the same order of magnitude. 


In the scaling limit the one-dimensional problem is integrable along the lines $g=0$ or $h=0$, and
conformally invariant when both these conditions are met.  Throughout this
work we use the standard conformal field theory normalization of the magnetization operator,
fixed by
\be
\chi_{1D}(\tau,x) = \la \sigma(\tau,x) \sigma(0) \ra =  \frac{(a_0/v)^{1/4}}{|r|^{1/4}},
\ |r|\rightarrow 0,
\label{norm}
\ee
where $r^2 = \tau^2 + (x/v)^2$, $\tau$ is the Matsubara time.

We first calculate the phase diagram of the model (\ref{ham}) plotted schematically in
Fig. \ref{phase1}(b).
To estimate  the critical temperature, we set $h=0$ and then look for
$\omega=0$ divergences  in the correlation function which signify a developing instability.

We begin by considering the line $g=0$ where the uncoupled chains are critical.
At finite temperature the spin-spin correlation function (\ref{norm}) becomes 
\be
\chi_{1D}(\tau,x) =
\left[\frac{(\pi Ta_0/v)^2}{\sinh (\pi T(x/v-\ri \tau)) \sinh (\pi T (x/v+\ri \tau))}
\right]^{1/8},
\ee
The Fourier transform (\ref{ftd}) gives the dynamic susceptibility \cite{sb83}
\be
\chi_{1D}(\omega=0,k=0) = \frac{a_0}{v} (2\pi Ta_0/v)^{-7/4} \sin\frac{\pi}{8} B^2(1/16,7/8)
\ee
where $B(x,y)=\Gamma(x)\Gamma(y)/ \Gamma(x+y)$ is the Beta function.
Using the RPA we extract the transition temperature:
\be
T_c/J_\| = 2.12 \left[\frac{z_{\perp}J_\perp}{J_\|}\right] ^{4/7}.\label{tcg0}
\ee

\begin{figure}
\begin{center}
\epsfig{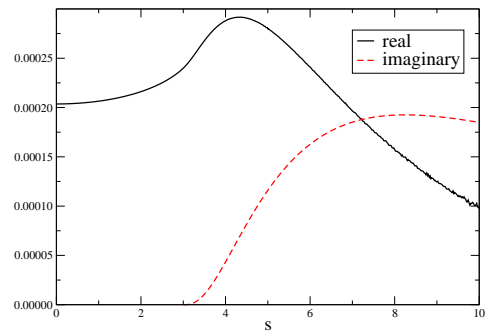}
\caption{The three particle contribution to the dynamical susceptibility for a single chain.}\label{p3}
\end{center}
\end{figure}

Now let us estimate the position of the Quantum Critical point on the $g$ axis. Again we use the RPA equation (\ref{rpa}) substituting in the expression for the dynamical spin susceptibility of the off-critical Ising model. According to  \cite{wmtb76}
\be
\chi_{1D}(r) = \frac{Z_0 (ma_0/v)^{1/4}}{\pi} K_0 (mr) + O ( e^{-3r} )
\ee
where $Z_0 = 1.8437$.  Hence for small $\omega$,
\bea
\chi_{1D}(\omega,k) &=& \frac{v}{a_0}\frac{Z_0 (ma_0/v)^{1/4}}{\omega^2-(vk)^2 - m^2} + \ldots 
\label{chi1}
\eea
where the dots indicate further terms in the form-factor expansion \cite{bb92}.  The next term, the three particle contribution can be represented as\cite{ess00}
\bea
&& \frac{\chi_3 (s^2)}{\chi_1(s=0)} = \nn
&& \frac{1}{3} \int\frac{dx}{4\pi}\int\frac{dy}{4\pi}
\frac{f(2x)f(y+x)f(y-x)}{ (s/m)^2 - \left[ 1 + 4\cosh x (\cosh x + \cosh y)\right]} \nn
&& f(x)  =  \tanh^2(x/2),
\eea
with $s^2 = \omega^2 - (vk)^2$ and is plotted in Fig. \ref{p3}.
The imaginary part of this term (which is the contribution to the structure factor) and all higher terms in the
series is $0$ below $\omega = 3m$, however all terms in the expansion contribute towards the real part at $\omega=0$.
Fortunately the contribution from higher order terms are negligible at small $s$ (see e.g. \cite{bn97}).  We see explicitly in this case that $\chi_3(s=0)/\chi_1(s=0)=0.0002$ so that (\ref{chi1}) is indeed a very good approximation for small $s$.

From Eq.\ \ref{rpa}, we can see that the QCP
where $T_c \rightarrow 0$ is given by the condition
\be
g_c \approx 1.42( z_{\perp}J_\perp /J_\|)^{4/7}.
\label{qcp1}
\ee
In the vicinity of QCP the low-energy behavior of the quantum Ising model is universal and falls in the universality class of the (d+2)-dimensional classical 
Ising model, where d is the number of transverse dimensions. Since d =2 corresponds to the upper critical dimension of that model, in three dimensions the fluctuations give only logarithmic corrections to RPA. 
The RPA expression for  the susceptibility at the QCP is 
\be
\chi (\omega,k; {\bf k_\perp} ) \sim \frac{1}{\omega^2 - (vk)^2 - ({\bf v}_\perp {\bf k}_\perp)^2}
\label{cps}
\ee
where ${\bf v}_\perp = (1/2)Z_0 g^{1/4} J_\|  {\mbox d}^2 J_\perp ({\bf k}_\perp=0)/{\mbox d}{\bf k}_\perp^2$.  


Near the critical point, we can examine the shape of the phase boundary.  For
low temperature, the correlation functions at finite temperature can be well approximated
\cite{qpt} by
\be
\chi_{1D} (\omega,k ) = \frac{Z_0 (m a_0)^{1/4}v^{1/4}}{(\omega+\ri/\tau_\psi)^2 - (v k)^2 - m^2}.\label{chiT}
\ee
where $\tau_\psi$ is the quantum dephasing time given by
\be
\frac{1}{\tau_\psi} = \frac{2T}{\pi} \re^{-m/T}.
\ee 
The RPA then gives the condition for a singularity at $\omega=0$, $k=0$ as
\be
m^2 + 1/\tau_\psi^2 = Z_0 J_\| J_\perp g^{1/4},
\ee
so the transition temperature in the vicinity of the critical point is approximately
\bea
T_c &=& \frac{m}{\ln(1/y) - \ln \ln(1/y)} \nn
y &=& \frac{\pi}{2}\left(Z_0  (J_\perp/J_\|) g^{-7/4} -1\right)^{1/2}.
\eea

For $g<0$, uncoupled chains are completely ordered at $T=0$, and at $T>0$ and $m$ not too small there is order on a length scale
\be
\xi_c = v (2mT/\pi)^{-1/2} e^{m/T}.
\ee
A crude estimate for the transition temperature can be is given by 
\be
\frac{J_\perp}{T_c} \left(\frac{\xi_c(T_c)}{v/m}\right)^2 \sim 1
\ee
which gives 
\be
T_c \approx \frac{2 m}{\ln(m/J_{\perp})}.
\ee


\begin{table}
\begin{center}
\begin{tabular}{|c|c|}\hline
 $m_i/m_1$ & $Z_i $ \\ \hline \hline
1.000 & 0.247159  \\ \hline
1.618 & 0.0690172 \\ \hline
1.989 & 0.0209579 \\ \hline
2.405 & 0.0122653 \\ \hline
2.956 & 0.0021898 \\ \hline
3.218 & 0.0011328 \\ \hline
3.891 & 0.0001623 \\ \hline
4.783 & 0.0000055 \\ \hline
\end{tabular}
\end{center}
\caption{The masses and weights of the particles of the one-dimensional quantum Ising model in a magnetic field (after \cite{dm95})  The first three particles lie below the incoherent continuum which begins at $\omega=2m_1$.}\label{mandz}
\end{table}

Now we go on to calculate the dispersion and spectrum within  the ordered phase, where we have to consider $h\ne 0$. Model (\ref{mf}) is exactly solvable at $g =0$ only \cite{zam88}. As we have mentioned,  the spectrum consists of eight particles which for completeness are listed in Table \ref{mandz}.   The contributions to the dynamical susceptibility for small $\omega$ decline quickly with the growth of the particle mass and the decrease in spectral weight of the mode. Therefore the magnetic susceptibility at $T =0$ can be well approximated by  keeping only the first three poles below the incoherent continuum
\be
\chi_{1D} (s) = \left(\frac{4m_1^2}{15\pi J_\| h}\right)^2 
\sum_{i=1}^{3} \frac{Z_i}{s^2-m_i^2}, \label{corr}
\ee
with $s$ defined as before.
In this model, the mass scale is given by\cite{fat94}
\bea
m_1(h)/J_\| &=& \alpha_1 (h/J_\|)^{8/15}, \nn
\alpha_1 &=& 4.40490858 
\eea
and the single-particle expectation value is
\bea
\la \sigma^z (0) \ra &=& \alpha_2 (h/J_\|)^{1/15}, \nn
\alpha_2 &=& 1.07496.
\eea
The self-consistency relations \ref{mfsc} therefore give
\bea
h/J_\| &=& \left[ \alpha_2 J_\perp (0)/J_| \right]^{15/14} \nn
m/J_\| &=& \alpha_1 \left[ \alpha_2 J_\perp (0)/J_\| \right]^{4/7}
\eea 

In the RPA (\ref{rpa}) the dispersion is given by the condition
\bea
\frac{1}{\tilde{J}_\perp} & = & \sum_i \frac{Z_i}{s^2-m_i^2} \nn
\tilde{J}_\perp & = & \left(\frac{4m_1^2}{15\pi h}\right)^2 J_\perp = \alpha_3 m_1^2 \frac{J_\perp ({\bf k_\perp})}{J_\perp (0)}
\eea
where the final relation is obtained from the self-consistent relations and $\alpha_3 = (4/15\pi)^2 \alpha_1^2/\alpha_2
= 0.130 $.  To second order in $J_\perp$ this is solved to give
\be
s^2/m_1^2 = m_i^2/m_1^2 + Z_i \tilde{J}_\perp \left[ 1 + \sum_{j\ne i} \frac{Z_j m_1^2 \tilde{J}_\perp}{m_i^2-m_j^2} \right].
\ee
In numerical values this gives for the first three modes
\be
(s/m)^2 \approx \left\{ \begin{array}{l}
1.0 + 0.032 \mbox{} J_\perp({\bf k_\perp})/J_\perp(0), \\
2.618 + 0.009 \mbox{} J_\perp({\bf k_\perp})/J_\perp(0). \\
3.956 + 0.003 \mbox{} J_\perp({\bf k_\perp})/J_\perp(0). \end{array} \right. \label{disp}
\ee 
The relative weights of each mode are not significantly changed from those for the pure one-dimensional case.

It is remarkable how weak the  dispersion in the perpendicular direction is, so the ordered
phase probably remains  very one-dimensional in character even when $J_\perp/J_\|$ is not very small.

We can obtain some results slightly away from the integrable lines.  In the neighborhood of the QCP where $h$ is
small, the dynamic susceptibility is given by (\ref{chi1})
with the mass replaced by \cite{dms96}
\be
m \rightarrow m[ 1+ (h/J_\|)^2 + O(h^4)].
\ee
For small $h$, we must also
have 
\be
\langle \sigma \rangle = \chi(\omega=0,k=0) h
\ee
which combined with the mean-field condition (\ref{mf}) gives
\bea
\frac{h}{J_\|} & =& \sqrt{\frac{1}{g\left[ J_\|/(Z_0 J_\perp (0) ) \right]^{4/7} } - 1} \\ \nn
m/J_\| &=& \left(Z_0 J_\perp(0)/J_\|\right)^{4/7}.
\eea
Here, the dispersion in the perpendicular direction is much stronger than in (\ref{disp}).  It is given
by (\ref{cps}) at the QCP; in the vicinity of $g_c$  the gap is given by $m \sim (g_c-g)^2$ in the ordered phase to be contrasted
with $m\sim (g-g_c)$ at $g>g_c$.

\begin{figure}
\begin{center}
\epsfig{file=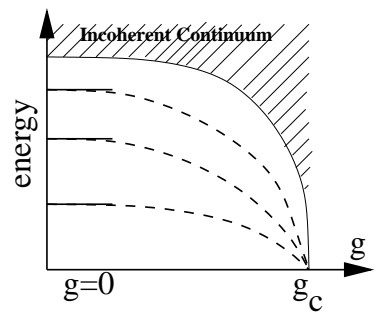,width=2in}
\caption{A schematic diagram showing what happens to the coherent modes as a function of $g$.  To the left of the figure, we have three coherent modes with very one-dimensional dispersions.  As we move to the right, the transverse dispersion grows and the overall mass scale decreases until all modes collapse at criticality.}\label{dispg}
\end{center}
\end{figure}

 Our results indicate that some of the beautiful physics of quantum Ising model in magnetic field with a hidden E$_8$ symmetry may be observed even in a realistic quasi-one-dimensional model in its ordered phase  far from the transition line (in the vicinity of point 0 on Fig. 1). At this point it may be possible to observe at least three  coherent peaks in the dynamical magnetic susceptibility whose relative strength is approximately $1:0.28:0.09$ (see Eqs.(\ref{corr}) and Table 1). At this point the spectrum is remarkably one-dimensional in character. We expect that when one moves along the $g$-axis of phase diagram of Fig. 1 towards the QCP, the excitation gaps will decrease and the transverse dispersion will grow. At the QCP the spectrum is three-dimensional gapless and the spin-spin correlation function is given by Eq.(\ref{cps}) with logarithmic corrections. These corrections will convert the pole into a continuum; at criticality   all excitation modes will collapse into it.  This is illustrated schematically in Fig. \ref{dispg}.

A good candidate experimental compound to see these effects is NaV$_2$O$_5$.  In \cite{mk00} it is shown that
the charge ordering at low temperatures is well described by the Hamiltonian (\ref{qim}) where the
$\sigma$ variables correspond to an isospin that describes which rung of a ladder the electron is sitting.  These isospin excitations should be experimentally probable with optical absorbtion \cite{mkk02}. 


The work is supported by the US DOE under
contract number DE-AC02-98 CH 10886 and EPSRC grant number 99307266.  STC would like to thank Fabian Essler
for many helpful discussions and Kasper Ericken for suggestions whilst preparing the manuscript.

\bibliographystyle{prsty}
\bibliography{../mybib.bib}

\end{document}